\documentstyle[aps]{revtex}

\input{psfig}
\begin{document}
\draft
\preprint{}

\title{Differential dispersion relations and elementary amplitudes
in a multiple diffraction model}

\author{A. F. Martini and M. J. Menon}

\address{Instituto de F\'{\i}sica `` Gleb Wataghin ''\\
Universidade Estadual de Campinas, Unicamp\\
13083-970 Campinas, SP, Brasil\\}

\author{ J. T. S. Paes\thanks{Deceased} and M. J. Silva Neto}

\address{Centro de Ci\^{e}ncias Exatas e Naturais - Departamento de 
F\'{\i}sica \\
Universidade Federal do Par\'{a} - Campus Universit\'{a}rio - 
Guam\'{a}\\
66075-900 - Bel\'{e}m, PA, Brasil}
\date{\today}
\maketitle

\begin{abstract}
We discuss the evaluation of the real part of the elementary amplitudes 
in the
context of a multiple diffraction model for $pp$ elastic scattering earlier
developed. The framework is based on the concepts of analyticity and 
polynomial boundedness, and the techniques of dispersion relations. 
Novel results concern the use of derivative dispersion relations at 
the elementary level (constituent-constituent interactions) and an 
optimization of these relations in terms of one free parameter. In addition to 
a theoretical improvement, we achieved a satisfactory description of the 
physical quantities.

\end{abstract}
\vskip 0.5truecm

\vskip 0.5truecm

\centerline{\bf I. INTRODUCTION}

\vskip 0.3truecm

In a previous paper, through a multiple diffraction model, a satisfactory 
description of $pp$ elastic scattering data at the highest energies was 
obtained \cite{martinimenon}. The approach is based on the Glauber 
formalism, which is characterized by two essential formulas \cite{glauber}:

\begin{equation}
F(s,q)=i\int bdbJ_{0}(qb)[1-e^{i\chi (s,b)}],
\end{equation}

\begin{equation}
\chi (s,b)=\int qdqJ_{0}(qb)G_{A}G_{B}f,
\end{equation}
where $F(s,q)$ is the hadronic amplitude, $G_{A,B}$ the hadronic form
factors and $f$, the elementary (constituent-constituent) amplitude. 
With inputs for 
$G_{A},G_{B}$ and $f$, physical quantities may be investigated,  
such as, the differential cross section

\begin{equation}
{d\sigma \over dq^{2}}=\pi \left| F(s,q)\right| ^{2},
\end{equation}
the total cross section

\begin{equation}
\sigma _{{\rm tot}}(s)=4\pi ImF(s,q=0),
\end{equation}
and the ratio of the forward real and imaginary parts of the hadronic
amplitude,

\begin{equation}
\rho (s)=\frac{ReF(s,q=0)}{ImF(s,q=0)}\text{ .}
\end{equation}

The model we referred to is characterized by the following 
parametrizations \cite{martinimenon}:

\begin{equation}
G_{A}=G_{B}\equiv G(s,q)={1 \over 1+q^{2}/\alpha ^{2}(s)} \
{1 \over 1+q^2/\beta ^{2}},
\end{equation}
\begin{equation}
f(s,q)=(\lambda +i)C(s)h(q),
\end{equation}
\begin{equation}
h(q)={1 - q^{2}/a^2 \over 1+q^4/a^4}\text{ ,}
\end{equation}
with two fixed parameters

\begin{equation}
a^{2}=8.20\ {\rm GeV^{2}}, \qquad
\beta ^{2}=1.80\
{\rm GeV^{2}}
\end{equation}
and three parameters depending on the energy,

\begin{equation}
\alpha ^{-2}(s)=2.57-0.217\ln (s)+0.0243\ln ^{2}(s) \quad 
({\rm GeV^{-2}}),
\end{equation}

\begin{equation}
C(s)=14.3-1.65\ln (s)+0.159\ln ^{2}(s) \quad ({\rm GeV^{-2}}),
\end{equation}

\begin{equation}
\lambda (s)=\frac{0.0695\ln (s/s_{0})}{1+0.118\ln (s/s_{0})+0.015\ln
^{2}(s/s_{0})}\text{ ,}
\end{equation}
where $s_{0}=400$ GeV.

With the exception of parametrization (12) for $\lambda(s)$, all the 
other choices are physically justified, as explained in [1] and 
references therein. Although Eq. (12) allows a good description of $pp$ 
experimental data and has also recently been used in analyses of 
$p$-nucleus collisions \cite{wibig}, it does not
have a full theoretical foundation, as recalled in what follows.

Physically, from assumption (7), $\lambda(s)$ plays at the elementary 
level the same role as $\rho(s)$ at the hadronic level, with the 
exception of the constancy in terms of momentum transfer:

\begin{equation}
\lambda(s) = {Ref(s,q=0) \over Imf(s,q=0)} = {Ref(s,q) \over 
Imf(s,q)} .
\end{equation}
As explained in [1], parametrization (12) was inferred from the 
observed similarities between $\rho(s)$ and $\lambda(s)$ and then 
from experimental information on $\rho(s)$, including $\overline{p} 
p$ scattering. For this reason, from the theoretical point of view, 
this parametrization may be seen as an {\it ad hoc} hypothesis.

In this paper we shall introduce novel parametrizations for 
$\lambda(s)$, but
now based on general principles of local quantum field theory QFT. 
The framework concerns
 the concepts of analyticity and polynomial boundedness and the 
techniques of dispersion relations.
 The paper is organized as follows. In Sec. II we review some 
theoretical
aspects concerning dispersion relations in the derivative form and 
apply these relations to the elementary amplitudes. With this we 
obtain an analytical expression for $\lambda(s)$, allowing a new 
determination of the real part of the hadronic amplitude. We shall 
also introduce an optimization of these relations
in terms of a free parameter. In Sec. III we discuss the results 
obtained and some aspects of the dispersion relations concerning 
both elementary and hadronic amplitudes, and also asymptotic limits. 
The conclusions and some final remarks are the content of Sec. IV.

\vskip 0.5truecm

\centerline{\bf II . DERIVATIVE DISPERSION RELATIONS
 AND ELEMENTARY AMPLITUDES}

\vskip 0.3truecm

Derivative dispersion relations were introduced nearly thirty 
years ago
\cite{gribov,bks}
and have, even recently, been used in the investigation of both 
$pp$ and $\overline p p$ scattering \cite{kawasaki,matthiae}. 
Its validity and practical applicability have also been extensively 
discussed \cite{kolar}. These relations may be extended to  an 
arbitrary number of subtractions, for both cross even and odd 
amplitudes and near the forward direction \cite{mmp}. Based on 
the general principles referred to before, it was shown that for 
an even amplitude the result for the first and second subtractions 
are the same and the leading term in the tangent series expansion 
reads \cite{mmp}

\begin{equation}
\frac{ReF_{+}(s,q)}{s^{\nu }}=\tan \left( \frac{\pi }{2}(\nu
-1)\right) \frac{Im F_{+}(s,q)}{s^{\nu }}+\frac{\pi }{2}\sec
^{2}\left( \frac{\pi }{2}(\nu -1)\right) \frac{d}{d\ln s}\left( \frac{
Im F_{+}(s,q)}{s^{\nu }}\right),
\end{equation}
where $\nu$ is a real free parameter constrained to the interval 
$0<\nu<2$. This result is valid for $s>>m^2$ and $q^2 \lesssim m^2$, 
where $m$ is the proton mass.

The smooth increase of the hadronic cross sections, roughly as 
$\ln^2s$, is compatible with the class of functions which verifies Eq.
(14). From the optical theorem, Eq. (4), this allows the simultaneous 
investigation of $\sigma_{{\rm tot}}(s)$ and $\rho(s)$  for both $pp$ 
and $\overline{p} p$ interactions \cite{matthiae}.
The conventional form widely applied to hadronic amplitudes corresponds 
to the particular case $\nu = 1$ and the forward direction:

\begin{equation}
{ReF_+(s,q=0) \over s} = {\pi \over 2} {d \over dlns} \left[{ImF_+(s,q=0) 
\over s}\right].
\end{equation}
Although all known uses of derivative dispersion relations in elastic and 
diffractive scattering concern hadronic amplitudes, in what follows we shall 
investigate its applicability at the elementary level.

In the model described in Sec. I, the imaginary part of the elementary 
amplitude
factorizes in the form

\begin{equation}
Imf(s,q) = C(s)h(q).
\end{equation}
We connect this quantity with the derivative dispersion relation (14) by 
defining \cite{caxambu97}

\begin{equation}
{ImF_+(s,q) \over s } \equiv Imf(s,q).
\end{equation}
From Eqs. (14), (16) and (17), we obtain an analytical expression for 
$\lambda(s)$ in terms of $C(s)$ and $\nu$

\begin{equation}
\lambda (s,\nu)=\frac{Ref(s,q)}{Imf(s,q)}=\tan \left( \frac{\pi }{2%
}(\nu -1)\right) +\frac{\pi }{2}\sec ^{2}\left( \frac{\pi }{2}(\nu
-1)\right) \left[ \frac{1}{C(s)}\frac{dC(s)}{d\ln s}+1-\nu \right] . 
\end{equation}
With parametrization (11) for $C(s)$ the above connection is valid, 
since the increase of this quantity with the energy follows a second 
degree polynomial in $\ln s $
(smooth increase). 
Differently from the {\it ad hoc} hypothesis (12), we now have a 
theoretically justified result for $\lambda(s)$, given in terms of the
parametrization for $C(s)$. For each $\nu$ value the model described in 
Sec. I leads to the physical quantities (3), (4), and (5).

As a first test we took the conventional value $\nu=1$. The corresponding
 behaviour of $\lambda(s,\nu =1)$ is shown in Fig. 1, together with 
the 
early parametrization (12). The results for the differential and
total cross sections are quite similar to the previous ones, 
presented in  \cite{martinimenon}. However, the predictions for 
$\rho(s)$ overestimate the experimental data, as can be seen in Fig. 2.

Based on this result we attempted to optimize the derivative relation 
(18) by letting free the parameter $\nu$. In this case we analyzed 
only the $\rho(s)$
data, where disagreements have been found. The best fit through the 
CERN-minut
\cite{minuit}
furnished
\begin{equation}
\nu = 1.25 \pm 0.01,
\end{equation}
with $\chi^2 = 7.76$ for $6$ degrees of freedom. The behaviour of 
$\lambda(s)$ is shown in Fig. 1 and the  prediction for the 
corresponding
$\rho(s)$ is displayed in Fig. 2. As before, the results for the 
differential and total cross sections are similar to the previous ones.

\vskip 0.5truecm

\centerline{\bf III. DISCUSSION}
\vskip 0.3truecm
A central point in this work concerns the introduction of Eq. (18) 
for the ratio of the real and imaginary parts of the elementary 
amplitude. In this section we discuss some consequences of this 
formula related with the predictions of physical quantities 
(descriptions of experimental data) and asymptotic behaviors.

\vskip 0.3truecm
\centerline{\bf A. Real part of the hadronic amplitude and differential 
cross sections}
\vskip 0.3truecm
An essential aspect of the multiple diffraction models is the connection
 between hadronic and elementary amplitudes. In applying the derivative 
dispersion relation at the elementary level we obtain both the real and 
imaginary parts of the hadronic amplitude. As ilustration
and for comparison, we calculate the real part of the hadronic amplitude 
through the parametrizations presented in Sec. I. The contribution of 
this part to the differential cross section, for $pp$ elastic scattering 
at $\sqrt s = 52.8$ GeV is shown in Fig. 3, together with the differential 
cross section data and for the three cases investigated, namely, 
$\lambda(s)$ from
Eq. (12) and $\lambda(s, \nu = 1.0)$, $\lambda(s, \nu=1.25)$ from Eq. (18).
We observe that in all cases the real part of the amplitude presents
two changes of signal and at the same positions. As showed in Ref. [1],
similar results are obtained, for example, through the Martin's formula
applied directly to the hadronic amplitude.

Since the imaginary part of the amplitude presents a zero at the dip
position, our results overestimate the experimental data
at this region, as can be seen in Fig. 3. This is the case even for the 
best result, corresponding to $\nu = 1.25$.
 However, as commented in Ref. [1], simultaneous and complete descriptions
 of total/differential cross sections and the $\rho$ parameter still 
remain an open problem in geometrical and multiple diffraction models. 
In our case, as in the previous approach, the limitations concern only 
the dip region at the highest ISR energies (see Fig. 4 in Ref. [1]). 
This seems to be consequence of the naive assumption of factorization 
of the imaginary part of the elementary amplitude, Eq. (16), leading to 
constant $\lambda$ in terms
of the momentum transfer. Recent results presented evidence for eikonal
zeros in the momentum transfer space \cite{carvalhomenon} and this may 
bring new insights on the search for a more suitable assumption. We are 
presently investigating this point.

\vskip 0.3truecm

\centerline{\bf B. Derivative dispersion relation and asymptotic limits}
\vskip 0.3truecm
A novel aspect of this work concerns the optimization of the derivative 
dispersion relation by letting free the parameter $\nu$. Contrary to some 
authors, which understand that ``...the choice of a $\nu$ different from 1 
has no practical advantage'' \cite{kolar}, we showed that in the context 
of our model, the best results were obtained with $\nu = 1.25$ and this 
demands further discussion.

We shall analyze some consequences of Eqs. (14) and (18) for $\lambda(s,\nu)$
 when
 $\nu \neq 1$ . It is
important to note that despite our analyses treat the elementary level,
similar results may be inferred at the hadronic level. In fact, as shown 
in Figs. 1 and 2 (and commented in Ref. [1]), $\lambda(s)$ and $\rho(s)$ are
strongly correlated: if $\lambda$ increases (decreases) also $\rho$ increases
(decreases), $\lambda=0$ at the same energy value where $\rho =0$, and the
same is true for the position of their maxima. Moreover, our parametrization 
for $C(s)$, Eq. (11), has the same functional form as the usual total cross 
sections parametrizations at sufficiently high energies. 

We first recall that the two-subtracted differential dispersion relation 
(14)
corresponds to the first term in a tangent series expansion, involving the 
derivative in the variable $\ln s$ \cite{mmp}. Since
this is an odd series and the parametrization for $C(s)$ is a second degree 
polynomial in $\ln s$, Eq. (14) is an ``exact'' result, that is, no other
terms in the expansion must be taken into account. Obviously the same is
true in the case of total cross sections parametrizations. 

According to
the approach of Ref. \cite{mmp} the $\nu$ parameter is constrained to the 
interval $(0,2)$ and formula (18) diverges at these extremes.
The behavior of $\lambda(s,\nu)$ for $\nu = 0.75,\ 1.00$, and $1.25$, is 
shown in Fig. 4, in
the wide energy interval $10^1 - 10^{15}$ GeV. Based on the strong 
correlation between $\rho(s)$ and $\lambda(s)$ (see Figs. 1 and 2) it is phenomenologically evident that, in order  to reproduce the experimental 
$\rho$ data,
we should expect values of $\nu$ near to 1. However a crucial point 
concerns the asymptotic behavior $\lambda(s\rightarrow \infty, \nu)$. 
From our parametrization for $C(s)$ the term involving this quantity
in Eq.  (18) vanishes at suficiently high energies. The asymptotic form
of $\lambda(\nu)$, from eq. (18), is displayed in Fig 5 in the above
interval of the parameter $\nu$. We see that $\nu = 1$ is a point of 
inflection
and the asymptotic values of $\lambda$ are positive for $\nu<1$ and 
negative
for $\nu>1$.

In the context of our model and parametrizations, the result $\nu=1.25$
leads to the asymptotic value $\lambda \sim -0.05$. From Fig. 2 it is 
evident that $\nu \leq 1$ cannot reproduce the experimental $\rho$ data. 
Therefore, a
second change of signal is predicted and the same is inferred for 
$\rho (s)$ at sufficiently high energies.

\vskip 0.5truecm

\centerline{\bf IV. CONCLUSIONS AND FINAL REMARKS}
\vskip 0.3truecm
In this work we discussed the evaluation of the real part of the 
elementary amplitude in the context of a multiple diffraction model 
earlier developed \cite{martinimenon}. The novel results are (a) 
application for
the first time of differential dispersion relations at the elementary 
level, allowing the introduction of a theoretically justified analytical 
parametrization for $\lambda(s)$, Eq. (18); (b) optimization of the 
derivative dispersion relation by letting free the parameter $\nu$.

Although we still have the parameter $\nu$ to be determined by fit, we 
stress that the behavior of $\rho(s)$ beyond the region with data 
available comes directly from the dispersion relation applied to the 
elementary amplitude. That is, this behavior is not imposed as in the 
previous parametrization for $\lambda(s)$, Eq. (12). From Fig. 2 we 
also observe a faster decrease of $\rho$ at the highest energies than 
in the previous approach.
In particular we predict for $pp$ elastic scattering: $\rho_{max} = 
0.14$ at
$\sqrt s \sim 580$ GeV and $\rho = 0.11$ at $\sqrt s = 16$ TeV (LHC).
We also presented indication that $\rho$ may become negative at 
sufficiently high energies: That is the case for $\nu > 1.0$.

This change of sign is obviously an unconventional result, bringing 
some
resemblance only with some kinds of Odderon analyses \cite{nico}. 
Concerning this point, it should be stressed that Eq. (14) is the general
formula for the derivative relation, as introduced by Bronzan, Kane and
Sukhatme \cite{bks}. The $\rho$ parameter and total cross section should 
then be correlated through Eq. (18). As we have shown, to assume 
$\nu = 1$
means to impose that $\rho$ goes asymptotically to zero through 
positive values, if in this region, $\sigma_{{\rm tot}} \sim \ln^2 s$. 
For this reason,
it should
be interesting to investigate the $pp$ and $\overline{p}p$ available 
data on
$\rho$ and $\sigma_{{\rm tot}}$, through Eq. (18), with adequate cross
symmetry conditions
and treating $\nu$ as a free parameter. This could bring new information 
on the asymptotic behavior of $\rho(s)$ and of total cross sections.

\vskip 0.5truecm 

\leftline{ACKNOWLEDGMENTS}

We are thankful to Capes and CNPq for financial support.

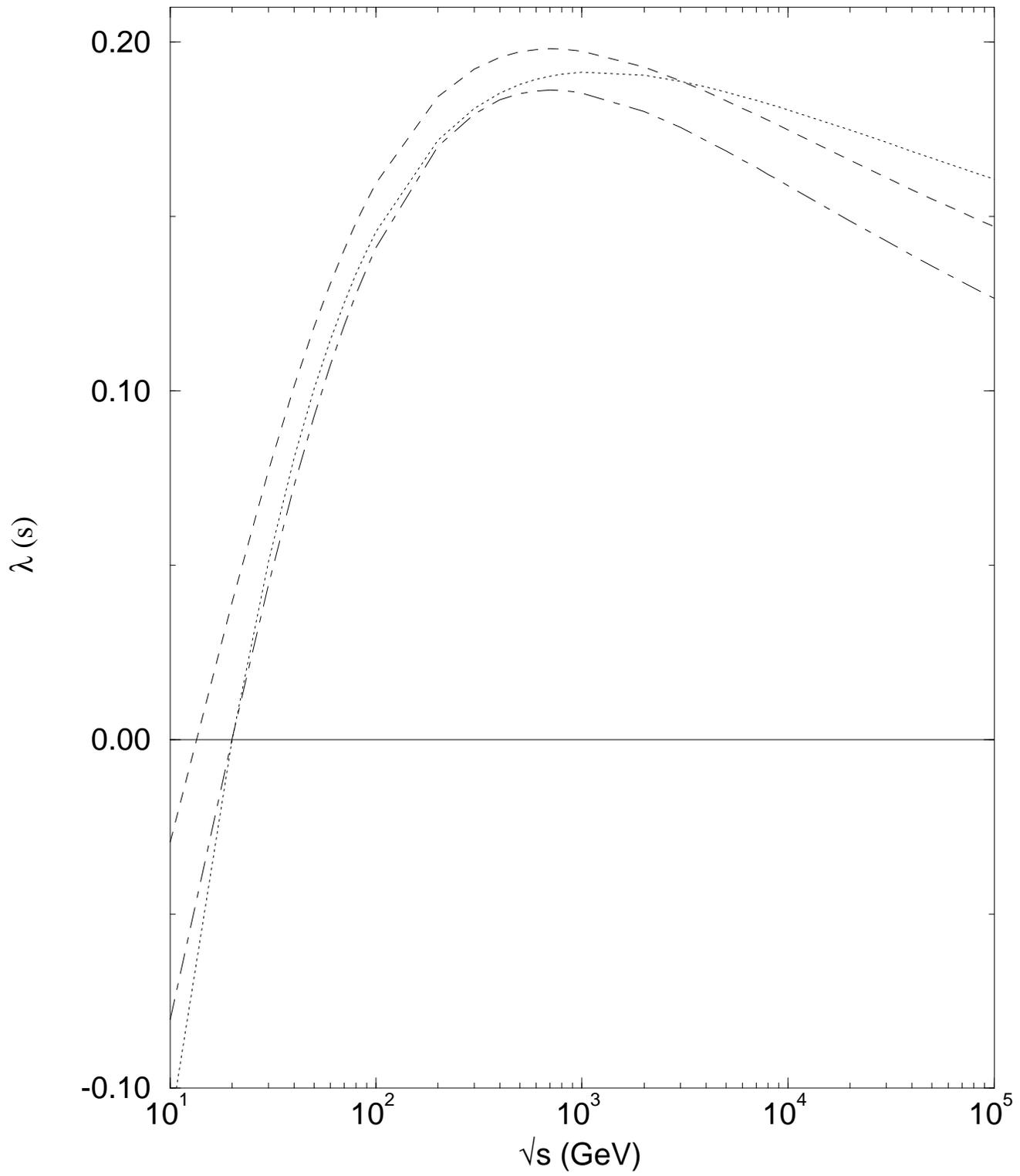
\begin{figure}
\caption{Ratio of the real to imaginary parts of the elementary 
amplitude,
$\protect\lambda(s)$: previous parametrization, Eq. (12) (dotted); 
results through the derivative dispersion relation, Eq. (18), with 
$\protect\nu = 1.0$
(dashed) and $\protect\nu = 1.25$ (dot-dashed).}
\label{fig1}
\end{figure}

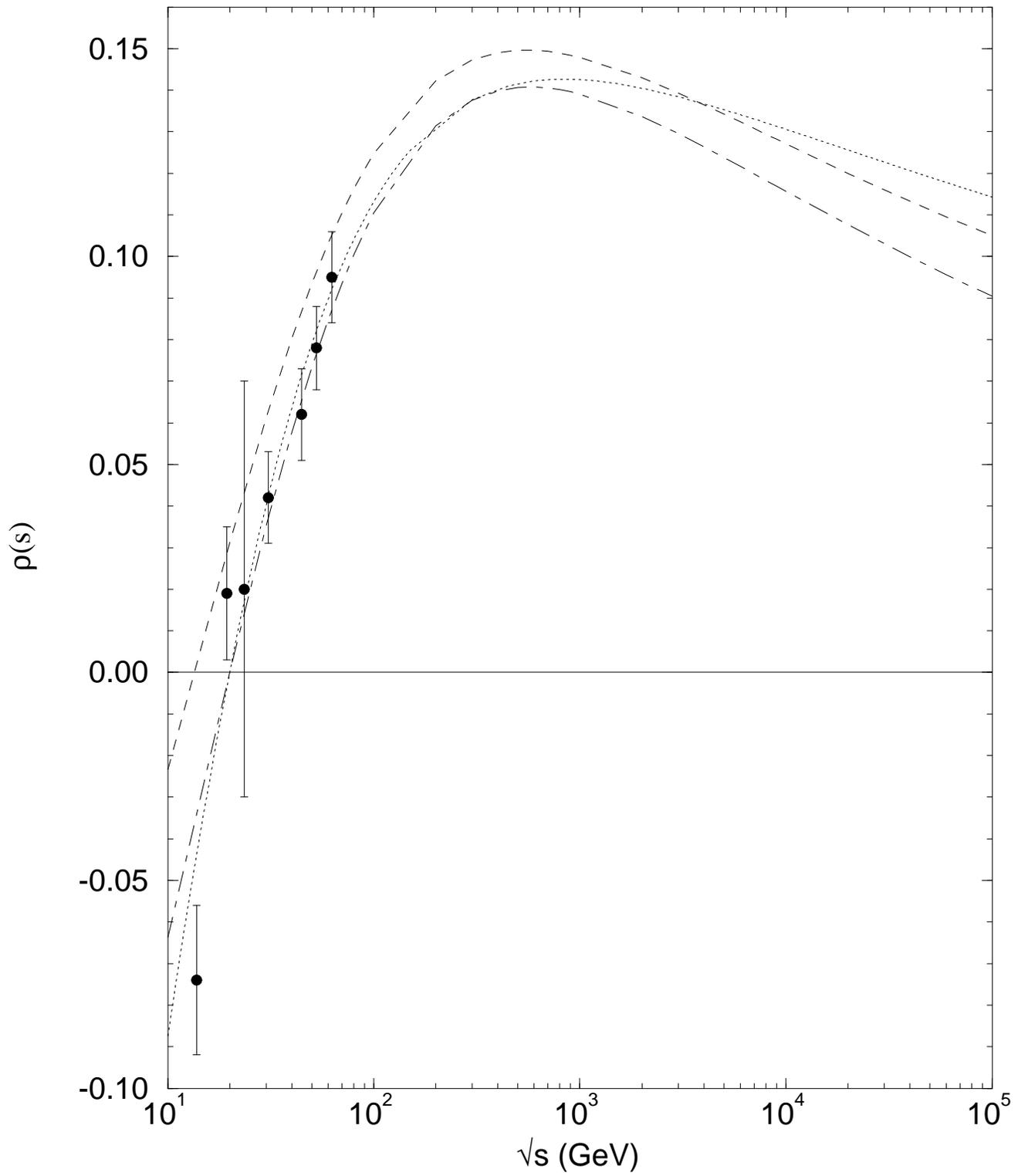
\begin{figure}
\caption{ Predictions for $\protect\rho(s)$ with $\protect\lambda(s)$ 
from Fig. 1 (same legend) and the model described in Sec. I, together 
with experimental data (See Ref. [1]).}
\label{fig2}
\end{figure}

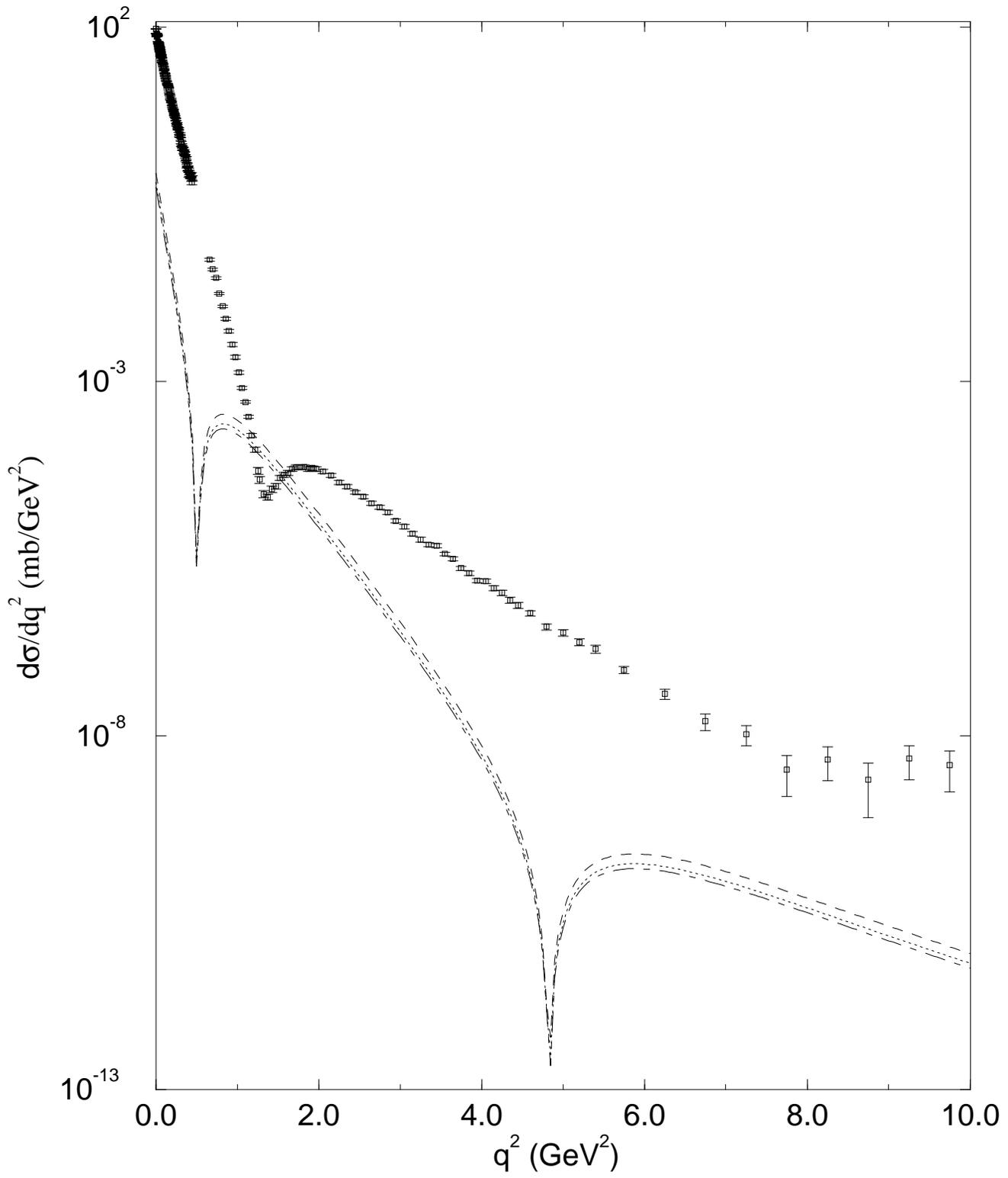
\begin{figure}
\caption{Contributions of the real part of the hadronic amplitude to 
the differential cross section, for $pp$ elastic scattering at 
$\protect\sqrt s = 52.8$ GeV and experimental data (see Ref. [1]). 
The curves were obtained with $\protect\lambda(s)$ from Fig. 1 (same 
legend) and the model described in Sec. I.}
\label{fig3}
\end{figure}

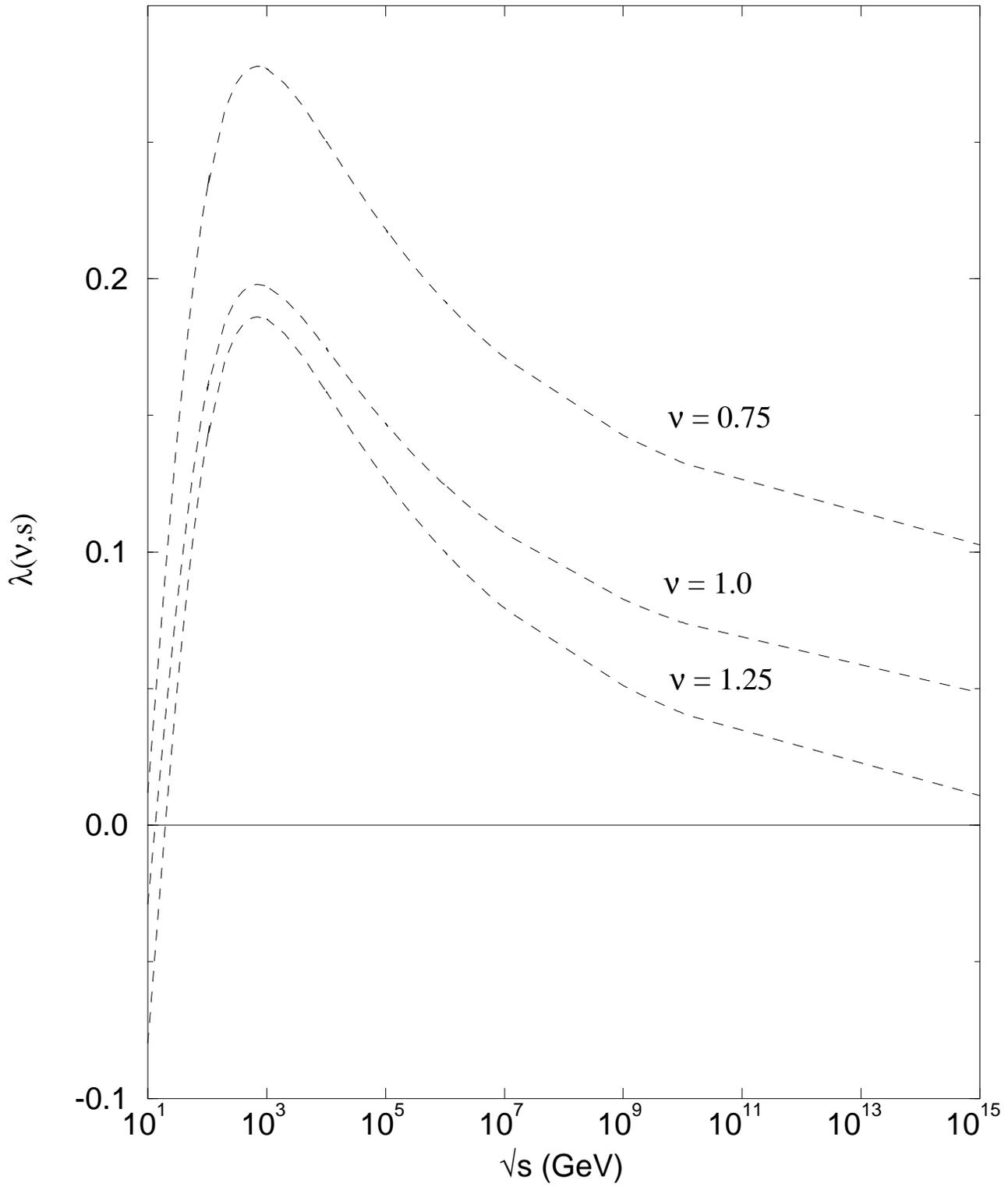
\begin{figure}
\caption{Results for $\protect\lambda(s, \nu)$ from Eq. (18) and 
parametrization (11) for $C(s)$.}
\label{fig4}
\end{figure}

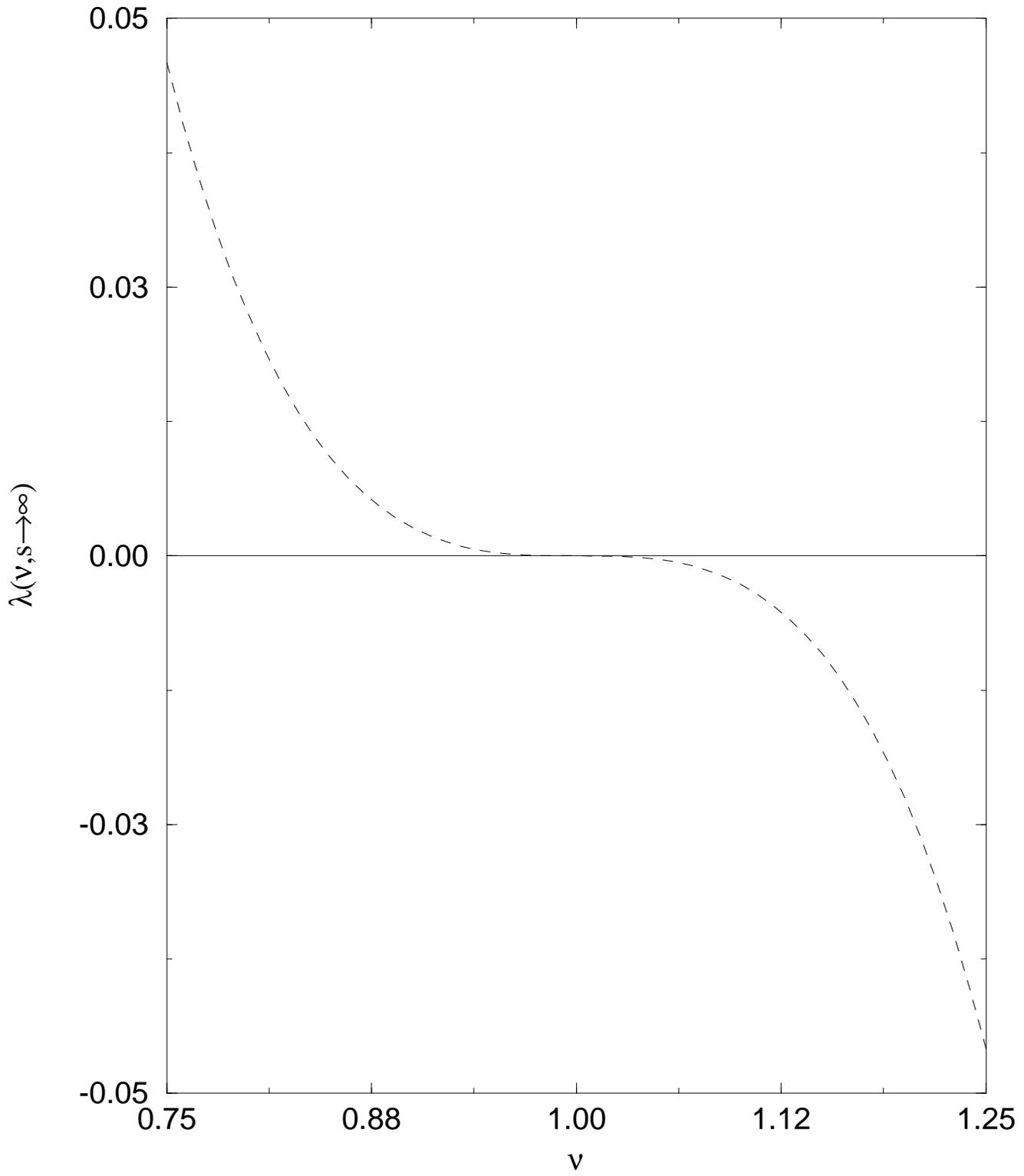
\begin{figure}
\caption{Asymptotic behavior of $\protect\lambda(\nu)$ from Eq. (18).}
\label{fig5}
\end{figure}

\end{document}